\documentclass[twocolumn,superscriptaddress,showpacs,aps,amsmath,amssymb,prb]{revtex4-1}

\usepackage{amsmath}
\usepackage{bm}
\usepackage{graphicx,color}
\usepackage{hyperref}

\newcommand{\nl}{\nonumber \\}

\newcommand{\Sec}[1]{Sec.\;\ref{#1}}

\newcommand{\be}{\begin{equation}}
\newcommand{\ee}{\end{equation}}
\newcommand{\bea}{\begin{eqnarray}}
\newcommand{\eea}{\end{eqnarray}}
\newcommand{\bal}{\begin{align}}
\newcommand{\eal}{\end{align}}
\newcommand{\bsube}{\begin{subequations}}
\newcommand{\esube}{\end{subequations}}
\newcommand{\Eq}[1]{Eq.\,(\ref{#1})}

\newcommand{\dg}{\dagger}
\newcommand{\la}{\langle}
\newcommand{\ra}{\rangle}

\begin{document}

 \title{Noise spectrum of a quantum dot-resonator lasing circuit}

\author{Jinshuang Jin }

\affiliation {Institut f\"ur Theoretische Festk\"orperphysik,
      Karlsruhe Institute of Technology, 76128 Karlsruhe, Germany}
\affiliation{ Department of Physics, Hangzhou Normal University,
  Hangzhou 310036, China}

\author{Michael Marthaler}
\affiliation {Institut f\"ur Theoretische Festk\"orperphysik,
      Karlsruhe Institute of Technology, 76128 Karlsruhe, Germany}
\author{Pei-Qing Jin}
\affiliation {Institut f\"ur Theoretische Festk\"orperphysik,
      Karlsruhe Institute of Technology, 76128 Karlsruhe, Germany}
\affiliation {Institute of Logistics Engineering, Shanghai Maritime University,
Shanghai 201306, China}
 \author{Dmitry Golubev}
\affiliation {Institute f\"ur Nanotechnologie, Karlsruhe Institute of Technology, 76021 Karlsruhe,
Germany}
 \author{Gerd Sch\"on}
\affiliation {Institut f\"ur Theoretische Festk\"orperphysik,
      Karlsruhe Institute of Technology, 76128 Karlsruhe, Germany}
\affiliation {Institute f\"ur Nanotechnologie, Karlsruhe Institute of Technology, 76021 Karlsruhe,
Germany}
\affiliation {DFG Center for Functional Nanostructures,
      Karlsruhe Institute of Technology, 76128 Karlsruhe, Germany}

\date{\today}

\begin{abstract}
Single-electron tunneling processes through a double quantum dot
can induce a lasing state in an electromagnetic resonator which is
coupled coherently to the dot system.
Here we study the noise properties of the transport current
in the lasing regime, i.e., both the zero-frequency shot noise as well as the
noise spectrum.
The former shows a remarkable super-Poissonian behavior
when the system approaches the lasing transition, but
sub-Poissonian behavior deep in the lasing state.
The noise spectrum contains
 information about the coherent dynamics of the coupled dot-resonator system.
It shows pronounced structures at frequencies matching that of the resonator
due to the excitation of photons.
For strong interdot Coulomb interaction we observe asymmetries in the auto-correlation noise spectra
of the left and right junctions, which we trace back to
asymmetries in the incoherent tunneling channels.

\end{abstract}

\maketitle

\section{Introduction}

A variety of fundamental quantum effects and phenomena
characteristic for cavity quantum electrodynamics (QED) have been demonstrated in
superconducting circuit QED. \cite{Wal04162,Chi04159,Bla04062320,Ili03097906}
The equivalent of single-atom lasing has been observed,
with frequencies in the few GHz range,  when a single Josephson charge qubit
is strongly coupled to a superconducting transmission
line resonator. \cite{Ast07588,Hau08037003} This progress stimulated
the study of a different circuit QED setup where the superconducting qubit
is replaced by a semiconductor double quantum dot
with discrete charge states.
Incoherent single-electron tunneling through the double dot 
sandwiched between two electrodes
can lead to a population inversion in the dot levels and, as a consequence, 
induce a lasing state in the resonator. \cite{Chi04042302,Jin11035322}
The potential advantages of quantum dots are
their high tunability, both of the
couplings and energy levels. 
 \cite{Fuj98932,Pet052180,Now071430}
In addition, larger frequencies are accessible since the restriction 
to frequencies below the superconducting gap is no longer needed.
Experimental progress has been made recently towards coupling semiconductor 
quantum dots to a GHz-frequency high quality
transmission line resonator. \cite{Fre12046807,Fre11262105,Del11256804}

The double quantum dot -- resonator circuit lasing setup 
differs from the more familiar
interband-transition semiconductor laser, \cite{Ben994756,Str11607}
where the cavity mode is coupled to the lowest quantum dot interband
transition, and which is driven by carrier injection in a p-n-junction or via optical pumping.
Since the circuit, considered here,
is driven by single-electron tunneling, the lasing state
correlates with electron transport properties. This fact allows probing
the former via a current measurement. \cite{Jin11035322}
Further information about the system is contained in the current fluctuations.
Due to the charge discreteness
the noise is shot noise. which has been studied extensively.
\cite{Her927061,Din9715521,Bla001,Naz03,Kie07206602,Gus06076605,Fuj061634}
For the double dot -- resonator lasing circuit it is therefore important to compare
the electron shot noise with the fluctuations of the photons in the resonator.

Although more difficult,
experimental progress has also been made towards
measuring the finite-frequency noise spectrum of electron transport. \cite{Ubb121}
It contains information about the full dynamics of
the system, including the relevant time scales that characterize the transport processes.
In the present work, we therefore investigate the frequency-dependent noise spectrum of
the transport current through the system in and near the lasing regime.
It shows pronounced characteristic signals at frequencies close to
the eigen-Rabi frequency of the coupled system or matching that of the resonator.

The present paper is organized as follows.
In \Sec{thmeth}, we introduce the model of
a quantum dot-resonator lasing circuit and the methods.
We extend the work of Ref.\,\onlinecite{Jin11035322},
where strong interdot Coulomb interaction was assumed,
to arbitrary strength interaction. \cite{Van031}
The method used for the calculation of the noise spectrum
is based on a master equation combined with the quantum regression theorem.
In \Sec{thsta}, the stationary properties of the resonator,
the average current, and the zero-frequency noise are studied.
The finite-frequency noise spectrum is evaluated in \Sec{thnoi} in the
 low- and high-frequency regimes,
both for strong and weak interdot Coulomb interaction.
We find characteristic symmetric and asymmetric  features in the
frequency-dependent noise spectrum.
We conclude with a discussion in \Sec{thsum}.

\section{Methodology}
\label{thmeth}
\subsection{Model}
\label{thCQD}

\begin{figure}[t]
 \centering
 \centerline{\includegraphics*[width=0.9\columnwidth,angle=0]{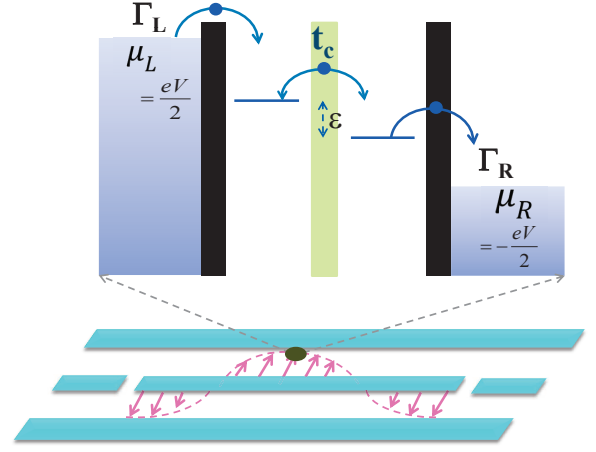}}
 \caption{(Color online) A double quantum dot--resonator
 lasing circuit. The dot is placed at a maximum of the electric field of the transmission line
  in order to maximize the dipole interaction.
  The population inversion in the dot levels, leading to the lasing state,
  is created by incoherent electron tunneling through the dots,
  driven by the bias voltage, which is assumed to be high, $eV=\mu_L-\mu_R\gg \omega_r$. }\label{dot}
 \end{figure}

We consider the electron transport setup schematically shown in Fig.\,\ref{dot},
where electrons tunnel through
a semiconducting double quantum dot coupled to a high-$Q$ electromagnetic
resonator such as a superconducting transmission line.
The Hamiltonian includes the interacting dot-resonator system,
$H_S=H_{\rm d}+H_{\rm r}+H_{\rm I}$, which is responsible for the coherent dynamics.
The  double dot is described by
 \begin{align}
 H_{\rm d}&=\sum_{j}\varepsilon_{ j}d^\dg_{j} d_{j}
 +Ud^\dg_{l} d_{l}d^\dg_{r} d_{r}+\frac{t_c}{2}\big(d^\dg_{l} d_{r}+d^\dg_{r} d_{l}\big),
 \label{Hdots}
 \end{align}
with $d_j^{\dg}$ being the electron creation
operators for the two levels in the dots $j$ ($j=l,r$)
with energies  $\varepsilon_{ j}$, separated by  $\varepsilon=\varepsilon_l-\varepsilon_r$,
which are coupled coherently with strength $t_c$. Both  $\varepsilon_{ j}$
and $t_c$ can be
 tuned by gate voltages. \cite{Oos98873, Har10195310,Hay03226804,Pet052180,Now071430}
The interdot Coulomb interaction is denoted by $U$.
The transmission line can be modeled as a harmonic oscillator,
 $H_{\rm r}=\omega_r a^\dg a$, with frequency $\omega_{\rm r}$
 and $a^\dg$ denoting the creation operator
 of photons in the resonator.
 The dipole moment induces an interaction
 between the resonator and the double dot,
 $ H_{\rm I}$, which will be specified below.

We further account for
electron tunneling between the dots and electrodes, $H_{\rm t}=\sum_{ k}  ( V_{L k }
c^{\dg}_{L k} d_{l }+ V_{R k }
c^{\dg}_{R k} d_{r }+\mbox{H.c.})$, with tunneling amplitudes $V_{\alpha k }$ (with $\alpha=L,R$).
The electrodes with  $H_{\rm b}
 =\sum_{\alpha k}\varepsilon_{\alpha k} c^{\dg}_{\alpha k}c_{\alpha k}$
 act as baths.
Here $c_{\alpha k}^{\dg}$  is the electron creation
operator for an electron state in the electrode $\alpha$.
Below, the tunneling between the electrodes and the dots is
assumed to be an incoherent process.

The double dot can be biased such that at most one electron occupies each dot. The two charge states $|L\ra$
and $|R\ra$ serve as basis of a charge qubit. \cite{Li01012302,Gur9715215}
In the present work, we consider two limits, (i) strong $U$ and (ii)
weak $U$, respectively. In case (i)  transport through
the double dots involves only
one extra third state, namely the empty-dot $|0\ra$, while
in case (ii) two extra states, $|0\ra$ and
the double occupation state $|2\ra\equiv|LR\ra$, are involved in the
transport. In both limits the  dipole interaction
 between the resonator and the double dot is,
 $ H_{\rm I} = \hbar g_0 (a^\dag+a^{}) \tau_z$,
 with Pauli matrices acting in the space of
 the two charge states, $\tau_z= |L\rangle\langle L|-|R\rangle\langle R|$.

 In the eigenbasis of the double dot and within rotating wave approximation,
 the Hamiltonian of the coupled dot-resonator system, for strong interdot Coulomb
 interaction, can be reduced to
 %
 \begin{eqnarray}\label{HRWA}
 H_{S}  = \frac{\hbar\omega_0}{2} \sigma_z
 + \hbar \omega^{}_{\rm r} a^\dag a
 + \hbar g (a^\dag \sigma_- +a^{} \sigma_+),
 \end{eqnarray}
while for weak interdot  interaction
an extra  term $U|2\ra \la 2|$ is to be included. In the restricted space of states we have
$d_l=|0\ra\la L|+|R\ra\la 2|$ and $d_r=|0\ra\la R|-|L\ra\la2|$,
and the Pauli matrix operates in the eigenbasis, i.e., $\sigma_z=|e\ra\la e|-
|g\ra\la g|$ with
 \begin{eqnarray}\label{eq_Eigenbasis_of_the_Dot}
 |e\rangle &=& \cos\left(\theta/2\right)|L\rangle + \sin\left(\theta/2\right)|R\rangle,
               \nonumber \\
 |g\rangle &=& \sin\left(\theta/2\right)|L\rangle - \cos\left(\theta/2\right)|R\rangle.
 \end{eqnarray}
 Here, we fix the zero energy level by
 $\varepsilon_l+\varepsilon_r=0$.
 The angle $\theta = \arctan(t_c/\varepsilon)$ characterizes the mixture of the pure charge states,
 the coupling strength is $g = g^{}_0 \sin\theta$, and
  $\omega_0 = \sqrt{\varepsilon^2+t^2_c}/\hbar$ denotes the level spacing of the
  two eigenstates.
 It can be tuned via gate voltages, which allows control of the
 detuning $\Delta = \omega_0-\omega_{\rm r}$
 from the resonator frequency.

\subsection{Master equation}
The dynamics of the coupled dot-resonator system, which is assumed
 to be weakly coupled to the electron reservoirs with smooth spectral density,
 can be described by
 a master equation for the reduced density matrix $\rho$
 in the Born-Markov approximation. \cite{Gar04,Car02}
Throughout this paper we consider low temperatures, $T =0$, with vanishing
 thermal photon number and excitation rates.
 Consequently, the master equation is
 \bsube\label{UME}
 \begin{align} \label{ME}
 \dot \rho &= -\frac{i}{\hbar}\left[H_{S}, \rho\right]
      + \mathcal L_{\rm L}\, \rho+\mathcal L_{\rm R}\, \rho
      +\mathcal L_{\rm r}\, \rho
 \equiv \mathcal L_{\rm tot} \, \rho,
 \end{align}
 where the dissipative dynamics is described by Lindblad operators of the form
 \begin{equation}\label{jump}
 {\cal L}_{i}\rho=\frac{\Gamma_i}{2}\left(2L_i\rho L_i^{\dag}-L_i^{\dag}L_i\rho-\rho L_i^{\dag}L_i\right).
 \end{equation}
 \esube
  The first two terms $ \mathcal L_{\rm L/R}$ account for the incoherent
  sequential tunneling between the
 electrodes and the dots
 with  $\Gamma_\alpha(\omega)=2\pi \sum_k |V_{\alpha k}|^2\delta(\omega-\varepsilon_{\alpha k})\equiv\Gamma_\alpha$. For the assumed high voltage and low
 temperature, i.e., in the absence of reverse tunneling processes, we
  have $L_{\rm L}=d^\dg_l$ and $L_{\rm R}=d_r$
 with tunneling rates $\Gamma_L$ and $\Gamma_R$, respectively.
 For the oscillator we take the standard decay term $L_{\rm r}=a$ with rate $\Gamma_{\rm r}=\kappa$.
Here, we ignore other dissipative effects,
such as relaxation and dephasing of the two charge states,
which were studied in Ref.\,\onlinecite{Jin11035322},
 since such effects only weakly
 affect the main points we wish to study.

From the definition $I_\alpha(t)\equiv -e\frac{d\la n_\alpha(t)\ra}{dt}$ with
$n_\alpha=\sum_k c^\dg_{\alpha k}c_{\alpha k}$, it is straightforward to obtain
the transport current from the electrodes to the dots, \cite{Lam08214302,Li05205304}
$I_\alpha(t)={\rm Tr}\big[ \hat I_{\rm \alpha}\rho(t)\big]$, with current operators given by
\bsube\label{curroperators}
\begin{align}
\hat I_{\rm L}\rho(t)&= \frac{e}{\hbar}\Gamma_{\rm L} d^\dg_l\rho(t) d_l,
\label{curroperatorL}
\\
 \hat I_R\rho(t)&= -\frac{e}{\hbar}\Gamma_{\rm R} d_r\rho(t)  d^\dg_r,
 \label{curroperatorR}
\end{align}
\esube
In the stationary limit, $t\rightarrow\infty$, the average current satisfies
 $I=\frac{1}{2}(I_L-I_R)=I_L=-I_R$,
consistent with  charge conservation.

\subsection{Current noise spectrum}

We consider the symmetrized current noise spectrum
\begin{align}
S(\omega)&={\cal F}\la \{\delta \hat I(t),\delta \hat I(0)\}\ra
\nl&\equiv
\int^\infty_{-\infty}dt\, e^{i\omega t}\la \{\delta \hat I(t),\delta \hat I(0)\}\ra
\nl&=2 \,{\rm Re}\big\{ \widetilde{G}_I(\omega)+\widetilde{G}_I(-\omega)\big\},
\end{align}
where $\delta\hat I(t)=\hat I(t)- I$ and $\widetilde{G}_I(\pm\omega)=\int^\infty_{0}dt\, e^{\pm i\omega t}G_I(t)$
with $G_I(t)=\la \delta\hat I(t) \delta\hat I(0)\ra$.
In Born-Markov approximation, the
current noise spectrum can be calculated via the widely used MacDonald's formula \cite{Mac62}
or the quantum regression theorem. \cite{Gar04} Since we already know the current operators,
as expressed
in \Eq{curroperators}, it is  more convenient to calculate the current correlation
function via the quantum regression theorem,
\be
G_I(t)={\rm Tr} \big[\hat I e^{{\cal L}_{\rm tot} t} \hat I \rho^{\rm st}\big]-I^2,
\ee
where $\rho^{\rm st}$ denotes the
steady-state density matrix.

According to the Ramo--Shockley
theorem, the measured
quantity in most experiments \cite{Bla001} is the total circuit current
$I(t)=aI_L(t)-bI_R(t)$, with
coefficients, $a+b=1$, depending on the
symmetry of the transport setup (e.g., the junction capacitances).
The circuit noise spectrum is thus composed of three components:
$S(\omega)=a^2 S_L(\omega)+b^2 S_R(\omega)-2ab S_{LR}(\omega)$,
\cite{Bla001,Eng04136602}
where $S_{\rm \alpha}(\omega)={\cal F}\la \{\delta \hat I_\alpha(t),\delta \hat I_\alpha(0)\}\ra$
are the auto-correlation noise spectra of the current from lead-$\alpha$,
 and $S_{\rm LR}(\omega)=
({\cal F}\la \{\delta \hat I_L(t),\delta \hat I_R(0)\}\ra+{\cal F}\la
\{\delta \hat I_L(t),\delta \hat I_R(0)\}\ra$)/2  is the current cross-correlation noise
spectrum between
 different leads.
Alternatively, in view of the charge conservation, i.e., $I_{\rm L}=I_{\rm R}+dQ/dt$,
where $Q$ is the charge on the central dots, the circuit noise spectrum can be
expressed as \cite{Moz02161313,Agu04206601,Luo07085325} $S(\omega)=a S_{\rm L}(\omega)+bS_{\rm R}(\omega)-ab\,S_{\rm C}(\omega)$
with $S_{\rm C}(\omega)\equiv{\cal F}\la \{\delta \dot{Q}(t),\delta \dot{Q}(0)\}\ra
=2S_{\rm L R}(\omega)+S_{\rm L}(\omega)+S_{\rm R}(\omega)$. \cite{Jin11053704}
Thus, from the behavior of the auto-correlation and cross-correlation noise
spectra, which will be studied in the following,
we can fully understand the circuit noise spectrum even including the charge
fluctuation spectrum in the central dots, $S_C(\omega)$.
At zero frequency, we have $S(0)=S_L(0)=S_R(0)=-S_{LR}(0)$
and $S_{\rm C}(0)=0$ due to current conservation in the steady-state.

\begin{figure}
\centerline{\includegraphics*[width=0.9\columnwidth,angle=0]{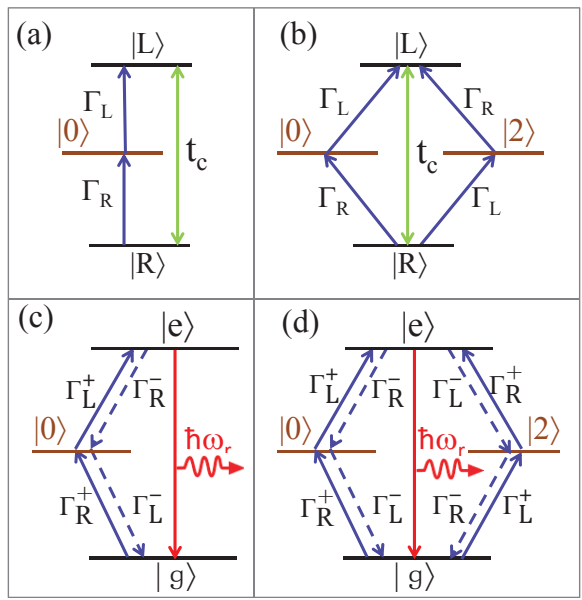}}
\caption{(Color online)  Incoherent electron tunneling induces transitions
 between different states in the dot. The upper panels, (a) and (b),
 and the lower ones, (c) and (d),
 describe the incoherent transitions in the
 dot-basis and eigen-basis (including the interaction with the resonator), respectively.
 Panels (a) and (c) corresponde to asymmetric transition channels for
 strong interdot Coulomb interaction, and (b) and (d) to symmetric transition channels for weak
 interdot Coulomb interaction. Furthermore, we have $\Gamma^+_{ \alpha}=\Gamma_\alpha \cos^2(\theta/2)$
 and $\Gamma^-_{ \alpha}=\Gamma_\alpha \sin^2(\theta/2)$ with $\alpha={\rm L, R}$. }
  \label{transition}
\end{figure}

\section{Stationary properties}
\label{thsta}
Let us first recall the parameter regime for which,
according to Ref.\,\onlinecite{Jin11035322}, lasing can be induced for the
present setup. \cite{Fre12046807,Fre11262105,Del11256804}
We consider the level spacing in the dots comparable to
the resonator frequency $\omega_r$ in the range of few GHZ,
and a high quality resonator with
$Q$ factor assumed to be $5\times10^4$, corresponding to a decay rate $\kappa=2\times 10^{-5}\omega_r$.
The coupling of dot and resonator, chosen as $g_0=10^{-3}\omega_r$, is strong enough
compared to the photon decay rate in the resonator,
  and we  assume weak
 incoherent tunneling  
 with
$\Gamma_{\rm L}=\Gamma_{\rm R}
=\Gamma=10^{-3}\omega_r$ to be a few MHz throughout of the paper,
 unless otherwise stated.

\begin{figure}
\centerline{\includegraphics*[width=1.0\columnwidth,angle=0]{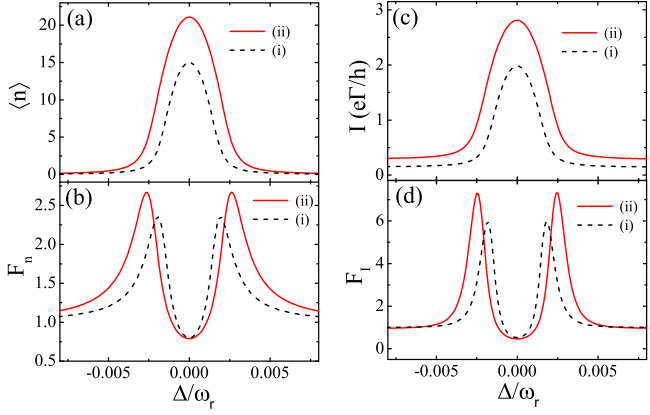}}
\caption{(Color online)
(a) Average photon number $\la n\ra$, (b) Fano factor $F_n=(\la n^2\ra-\la n\ra^2)/\la n\ra^2$ of
photons in the resonator, (c) average current $I$, and (d) Fano factor of the current $F_I=S(0)/2I$, as function of detuning for weak Coulomb interaction
(red solid line) and
strong interaction (black dashed line).
Throughout this paper we choose the tunneling rate $\Gamma=10^{-3}\omega_r$ and
interdot coupling strength $t_c=0.3\omega_r$.
} \label{stationary}
\end{figure}

A crucial prerequisite for lasing  is a
pumping mechanism, \cite{Mu925944,Ast07588} involving a third state,
which creates a population inversion in the two-level system.
In Ref.\,\onlinecite{Jin11035322}, the empty state $|0\ra$
in the double dot was considered as the single third state
under the assumption of strong charging energy, $\varepsilon_j+U>\mu_L>\varepsilon_j>\mu_R$.
This limit, which we call case (i), is sketched in Fig.\,\ref{transition} (a).
On the other hand, the interdot Coulomb interaction may also be
weaker compared to the level spacing of the charge states.
In the tunneling regime we have $\mu_L>\varepsilon_j,
 \varepsilon_j+U>\mu_R$. This limit, called case (ii),
where two extra states are involved in the incoherent tunneling, is
illustrated in Fig.\,\ref{transition} (b).
The question arises, which case is better for lasing.

Let us first consider the key factor for lasing, i.e., the population inversion
defined by
$\tau_{0}=\left(\rho^{\rm st}_e-\rho^{\rm st}_g \right)/\left( \rho^{\rm st}_e+\rho^{\rm st}_g\right)$,
with $\rho^{\rm st}_i=\sum_n\la i,n|\rho^{\rm st}|i,n\ra$ being
the stationary population of the
state of the dots ($i=e,g$). Explicitly, we find
\begin{align}
\tau_{0}&=\left\{\begin{array}{ll}
\frac{(\Gamma^2_{\rm R}/\omega^2_0 +4)\cos\theta}{\Gamma^2_{\rm R}/\omega^2_0+3 +\cos2\theta};
 &\qquad\quad\text{ for case (i)}
\\
\\
\frac{(\Gamma^2_0/\omega^2_0 +4)\cos\theta}{\Gamma^2_0/\omega^2_0+3 +\cos2\theta}; &\qquad\quad
\text{ for case (ii)}
\end{array}\right.,
\end{align}
with $\Gamma_0=\Gamma_{\rm L}+\Gamma_{\rm R}$.
The population inversion does not depend on $\Gamma_{\rm L}$ for case (i).
But it depends on both tunneling rates for case (ii),
suggesting that in this case the population inversion is driven
by transitions from $|R\ra$ to both
extra states $|0\ra$ and $|2\ra$. See Fig.\,\ref{transition} (a) and  (b)
for case (i) and (ii), respectively.
Although in general, an additional incoherent tunneling channel
reduces the population inversion slightly,
for the parameters studied in the present work, i.e., $\Gamma\ll\omega_r$,
it approaches the same
value  for both cases (i) and  (ii), $\tau_{0}\approx 4\cos\theta/(3 +\cos2\theta)$,
which reaches a maximum, $\tau_0\rightarrow 1$, for $\theta\rightarrow \pi/2$.
To balance the effective dot-resonator coupling $g=g_0 \sin\theta$ and the population inversion $\tau_0$,
following the consideration in Ref.\,\onlinecite{Jin11035322}, we set the interdot
coupling strength $t_c=0.3\omega_r$ throughout this work.

The properties of the resonator can be characterized by the
average  number of photons $\la n\ra$ and the Fano
factor $F_n\equiv\left(\la n^2\ra-\la n\ra^2\right)/\la n\ra^2$
describing their fluctuations. \cite{Str11607}
When reducing the detuning between dot and resonator from large values to zero,
we observe that 
the system undergoes a transition from the nonlasing regime, where $\la n\ra<1$
and $F_n=\la n\ra+1$, to a lasing state with a sharp increase
in the photon number. Before we reach the lasing state the photon number distribution has a thermal shape, which explains the
value of the Fano-Factor.
At the transition to the lasing regime the amplitude fluctuations lead to a peak in the Fano factor,
 as shown in Fig.\,\ref{stationary} (b).
In the lasing state the photon number is saturated, and the Fano factor drops to $F_n<1$, indicating a
squeezed photon number distribution in the resonator. Interestingly, the
average photon number in the lasing regime,
 as well as the corresponding peak in the Fano factor at the lasing transition
are larger for weak interdot interaction, case (ii), than for strong one, case (i).
Approximately, we obtain the average photon number \cite{Mar09} for case (ii)
\be\label{avern}
\la n \ra\simeq \frac{\Gamma\cos\theta}{2\kappa}-\frac{\Gamma^2+4\Delta^2}{8g^2}.
\ee
Compared to case (i), where \cite{Jin11035322} $\la n \ra_{i}\simeq\frac{\Gamma\cos\theta}{3\kappa}-\frac{\Gamma^2+4\Delta^2}{96g^2}(7+\cos\theta)$,
we find an increase to $\la n\ra _{ii}\approx \la n\ra _{i} +\frac{\Gamma}{6\kappa}$,
showing that case (ii) with four levels is more suited for lasing. The difference is
due to the existence of one more incoherent tunneling channel,
driven as illustrated in Fig.\,\ref{transition} (b) and (d).

Since photons in the resonator are excited by the
  incoherent tunneling between the dot and the electrodes,
  the lasing state closely correlates with the transport current.
  The current can be expressed approximately (for $\kappa\ll \Gamma$ and small $\theta$)
   for case (i) as \cite{Jin11035322}
\be
I(\Delta)\simeq e\Gamma \sum_{n=0} P(n)\frac{(n+1)}{3(n+1)+(\Gamma^2+4\Delta^2)/4g^2}
\ee
  with $P(n)\simeq (\Gamma/\kappa) P(0) \prod^n_{l=1}[3l+(\Gamma^2+4\Delta^2)/4g^2]^{-1}$
  being the probability of $n$ photons in the resonator
  (in Ref.\,\onlinecite{Jin11035322} a factor $\frac{1}{2}$ was missing).
  As shown in Fig.\,\ref{stationary} (c), the transport current as function of the detuning follows
  closely the behavior of the average photon number. 
  Similarly, the corresponding transport current for case (ii) is obtained  as,
   \be
  I(\Delta)\simeq e\Gamma \sum_{n=0} P(n)\frac{(n+1)}{2(n+1)+(\Gamma^2+4\Delta^2)/4g^2},
  \ee
  where $P(n)\simeq (\Gamma/\kappa) P(0) \prod^n_{l=1}[2l+(\Gamma^2+4\Delta^2)/4g^2]^{-1}$.
  Both the average photon number of the resonator as well as
  the current are larger for case (ii) than for case (i).

  As had been pointed out in Ref.\,\onlinecite{Har08024513},
  for a superconducting single-electron transistor
  (SSET) coupled to a resonator,
  the noise spectra of the fluctuations of the photons are correlated with the zero-frequency shot noise
  of the current. This fact is illustrated for
  the Fano factor $F_I=S(0)/2I$ in Fig.\,\ref{stationary} (d).
  For strong detuning in the nonlasing regime, where the dots effectively
  do not interact with the resonator, the shot noise shows a Poissonian
  distribution, i.e., $F_I\simeq 1$.
  Near the lasing transition the shot noise is enhanced strongly with
 a super-Poissonian distribution.
 Compared to the Fano factor of the photons, the signal in the shot noise is
   stronger with a narrower transition window and sharper peak.
  In the lasing state, where the
  photons are saturated and the transport current reaches the maximum value,
  we find sub-Poissonian current noise, $F_I\simeq 0.5$,
  while the photon Fano factor $F_n$ describes a squeezed state of the radiation field in
  this nonclassical regime, differing from a conventional coherent state with Poissonian distribution.
  The cross-correlation noise (not displayed in the figures)
 shows a similar behavior with the opposite sign due to
 the relation of $S_L(0)=S_R(0)=-S_{LR}(0)$.

\begin{figure}
\centering
\centerline{\includegraphics*[width=1.0\columnwidth,angle=0]{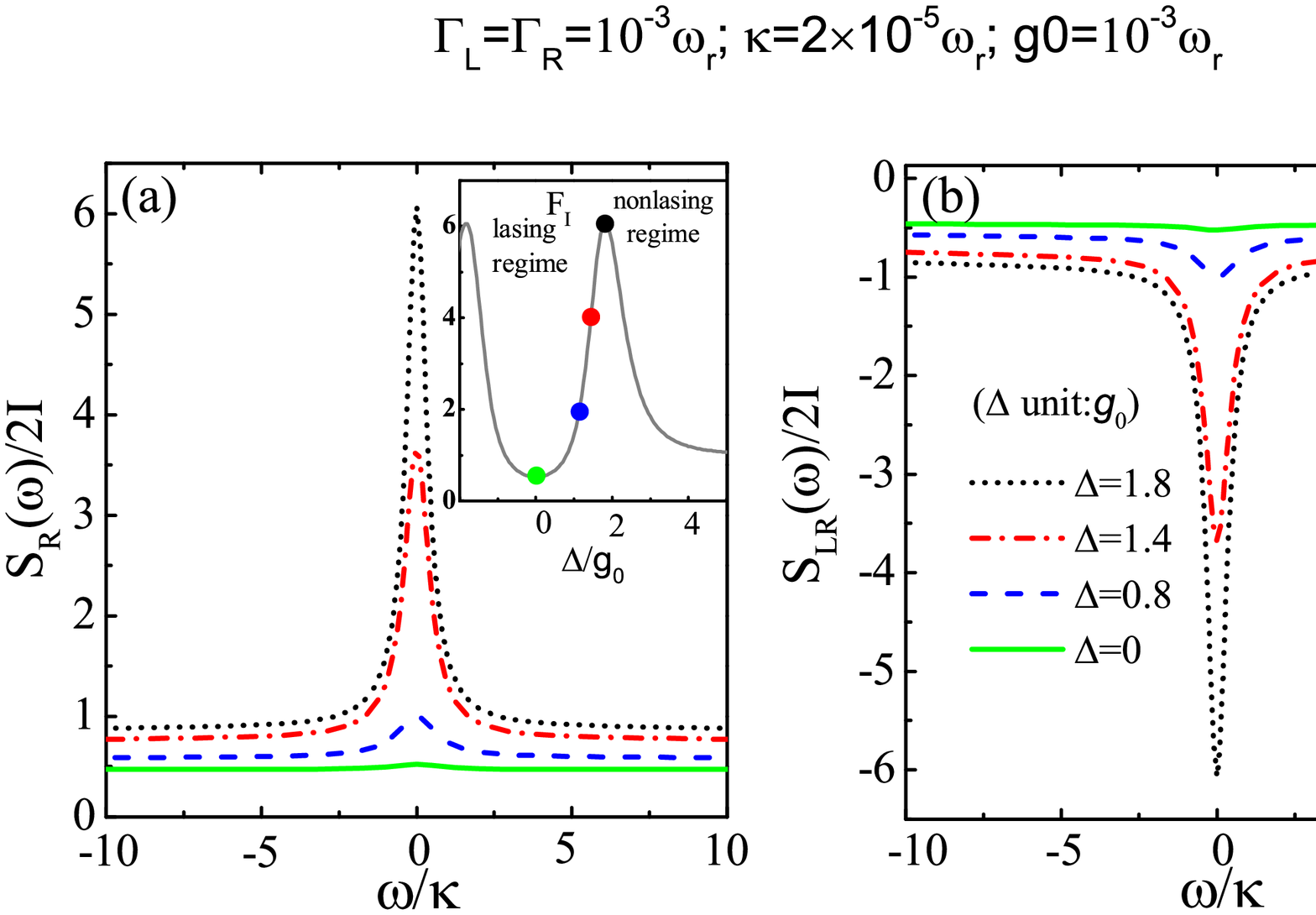}}
\caption{ (Color online)
The noise spectra in the low-frequency regime for strong interdot Coulomb interaction.
(a) Auto-correlation $S_L(\omega)=S_R(\omega)$. (b) Cross-correlation $S_{LR}(\omega)$.
Different colors of the plotted noise spectra
refer to different values of the detuning, as denoted in the inset of (a) by color circles.
The other parameters are the same as in Fig.\,\ref{stationary}.
} \label{Sw0-Uinf}
\end{figure}

\section{Noise spectrum}
\label{thnoi}

Since in the nonlasing regime the noise spectrum
displays only trivial features,
we focus in the following on
 the finite-frequency noise spectra
in the lasing regime and at the lasing transition,
as shown in the inset of
Fig.\,\ref{Sw0-Uinf}(a).
For tunneling dissipative operators ${\cal L}_L$ and ${\cal L}_R$
as defined after \Eq{jump} 
it has been demonstrated \cite{Ema07161404} that all correlation functions
can be expanded
in terms of the eigenvalues $\lambda_k$ of ${\cal L}_{\rm tot}$ and the
coefficients $c_k=[\hat V^{-1}\hat I_\alpha \hat V]_{kk}$. Here $\hat V$
is built from the eigenvectors of ${\cal L}_{\rm tot}$, and $\hat I_\alpha$ is the
current operator described in \Eq{curroperators}.
E.g., we have 
\begin{align}\label{Sw-eq}
\frac{S_\alpha(\omega)}{2 I}&=1-2\sum_k \frac{{\rm Re}(c_k){\rm Re}(\lambda_k)
+{\rm Im}(c_k)[\omega+{\rm Im}(\lambda_k)]}
{[\omega+{\rm Im}(\lambda_k)]^2+[{\rm Re}(\lambda_k)]^2},
\end{align}
where the imaginary part ${\rm Im}(\lambda_k)$ and real part
${\rm Re}(\lambda_k)$ represent
 the coherent and dissipative dynamics, respectively.
The coherent dynamics follows from the Jaynes-Cummings Hamiltonian, \Eq{HRWA},
with eigenstates \cite{Bla04062320,Har92}
\begin{align}
|+,n\ra&=\cos\theta_n|e,n\ra +\sin\theta_n|g,n+1\ra,
\\
|-,n\ra&=\sin\theta_n|e,n\ra -\cos\theta_n|g,n+1\ra,
\end{align}
and eigenergies
\be\label{eigenE}
E_{\pm,n}=(n+1)\omega_r\pm\frac{1}{2}\sqrt{4g^2(n+1)+\Delta^2}.
\ee
with $\theta_n=\frac{1}{2}\tan^{-1}\left(\frac{2g\sqrt{n+1}}{\Delta}\right)$.
The typical signal in the noise spectrum is dominated by
these eigenenergies, while the  linewidth of the signal follows from the jump operators in \Eq{jump}.

\begin{figure}
\centering
\centerline{\includegraphics*[width=1.0\columnwidth,angle=0]{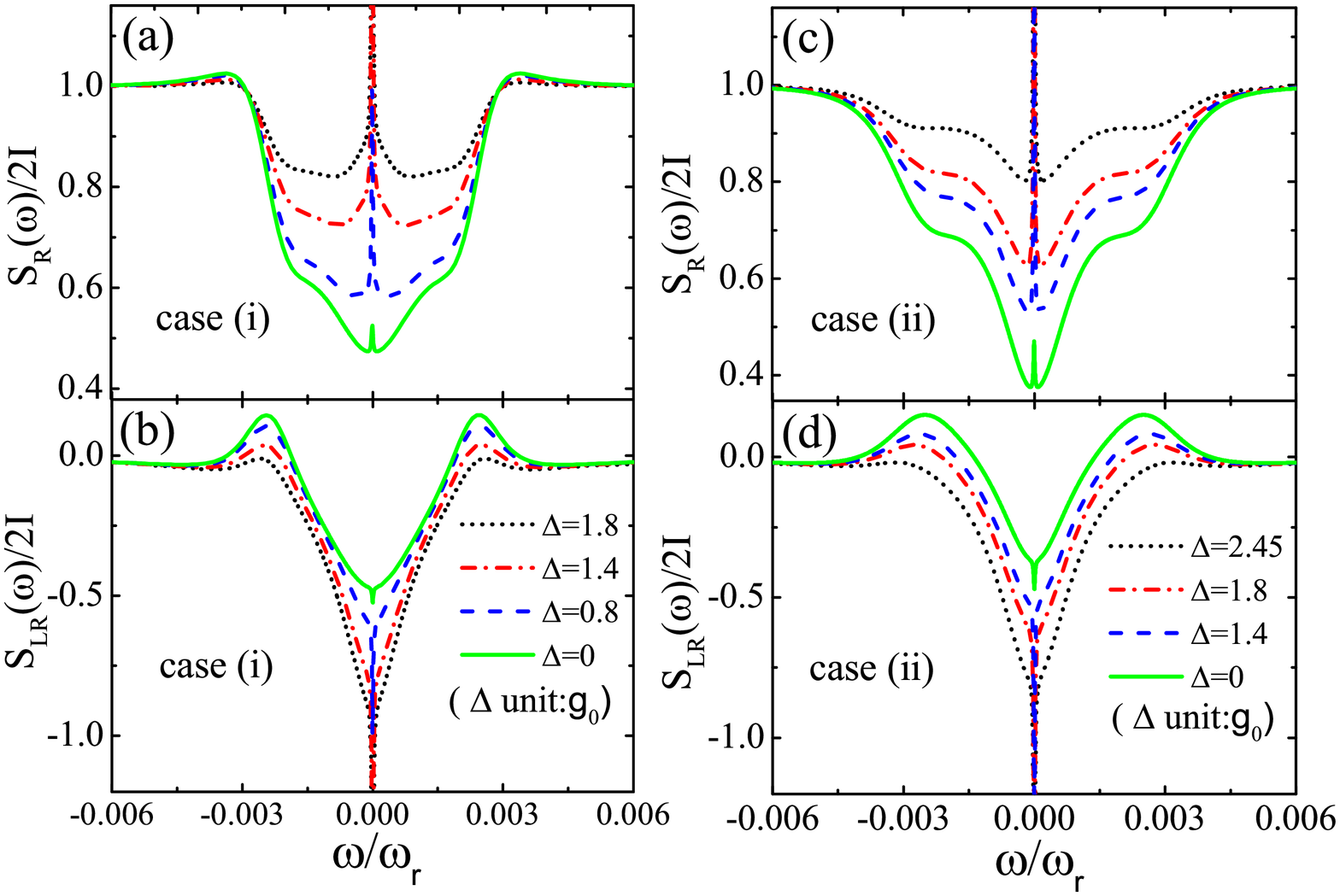}}
\caption{ (Color online)
The low-frequency auto-correlation and cross-correlation noise spectrum
for different values of the detuning
in the lasing regime.
Panel (a) and (b) are for strong interdot Coulomb interaction, and
(c) and (d) for weak interdot Coulomb interaction.
The other parameters are the same as in Fig.\,\ref{stationary}.
} \label{Sw-low}
\end{figure}

\subsection{Low-frequency regime}\label{subsec_lowfrequency}

Let us first consider the low frequency regime around $\omega\sim 0$
displayed in Fig.\,\ref{Sw0-Uinf}.
We find a zero-frequency peak and dip in the auto- and
cross-correlation noise spectra, respectively.
Both decrease and finally disappear
when one approaches the lasing state.
The height of the zero-frequency peak as function of a detuning
is shown in Fig.\,\ref{stationary} (d).
Since in the absence of a resonator we have
 $S_{\alpha}(\omega\approx 0)/2I = -S_{LR}(\omega\approx 0)/2I\simeq 1$,
the peak/dip feature at zero-frequency in the noise spectra must be the
effect of the resonator.

The noise spectra in Fig.\,\ref{Sw0-Uinf} have a Lorentzian shape with linewidth $\gamma_0\sim\kappa$,
determined by the emission spectrum of the photons. \cite{Car99}
In the regime around zero-frequency, corresponding to the long-time limit,
the noise spectra are determined
by the single minimum eigenvalue $\lambda_{\rm min}$ with
real part dominated by the weakest decay rate, i.e., $\kappa$.
For weak interdot Coulomb interaction, where we have to account for one more incoherent
tunneling channel (see Fig.\,\ref{transition} (b) and (d)) the low-frequency noise
spectra display a similar behavior as in Fig.\,\ref{Sw0-Uinf} (d),
except for the enhancement of the zero-frequency peak as shown in Fig.\,\ref{stationary}.
It is worth noting that in this low-frequency regime, the relation $S_L(\omega\sim 0)
=S_R(\omega\sim 0)=-S_{LR}(\omega\sim 0)$ is still satisfied. However,
as we will show below, the cross-correlation noise changes sign
beyond the low-frequency regime. 

\begin{figure}
\centering
\centerline{\includegraphics*[width=1.05\columnwidth,angle=0]{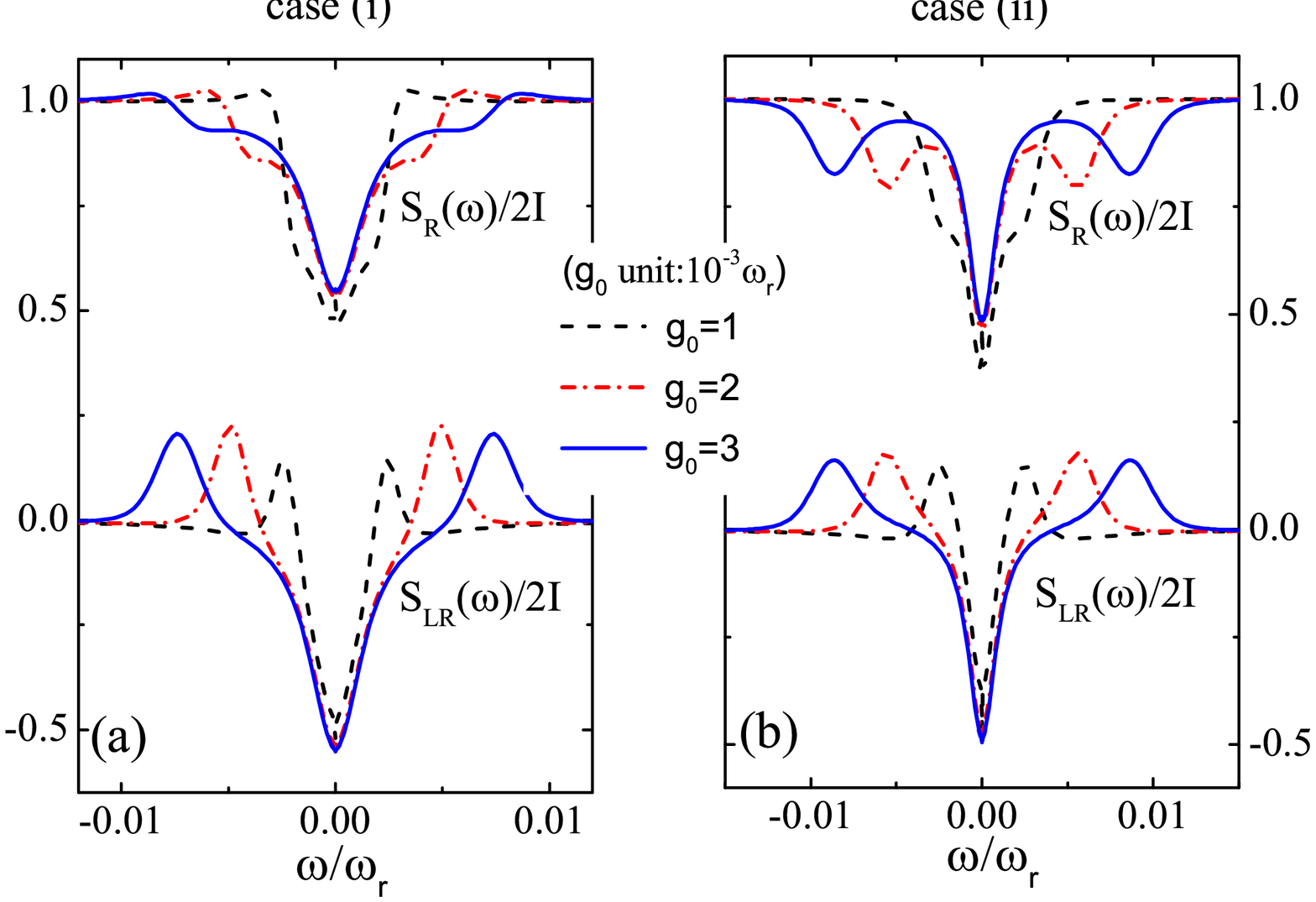}}
\caption{ (Color online)
The low-frequency noise spectra for different dot-resonator coupling strength
in the lasing state at $\Delta=0$,
(a) for strong interdot Coulomb interaction, and
(b) for weak interdot Coulomb interaction.
The other parameters are the same as in Fig.\,\ref{stationary}.
} \label{Slw-Gg}
\end{figure}

At higher frequencies but still within the range $|\omega|<\omega_r$,
the spectra are no longer Lorentzian due to the contributions from several
 $\lambda_k$ in \Eq{Sw-eq}.
We find characteristic features
showing a step and peak in the auto- and cross-correlation noise spectra, respectively,
as shown in Fig.\,\ref{Sw-low}. The position of the step/peak is nearly
independent of the detuning, while the magnitude is sensitive to it. 
With increasing dot-resonator interaction, both
the step and peak are shifted as shown in Fig.\,\ref{Slw-Gg}.
These characteristics are a consequence of the coherent dynamics of the
coupled dot-resonator system. The step/peak
occurs at $\omega=\delta E$, where
$\delta E=|E_{+,\langle n\rangle}-E_{-,\langle n\rangle }|=\sqrt{4g^2(\langle n\rangle +1)+\Delta^2}\approx 2g\big(\la n\ra+1\big)$
is the  Rabi frequency
corresponing to the photon number $\langle n \rangle$.
As expected this coherent signal of the step/peak becomes
weak and even disappears with increasing incoherent tunneling rate $\Gamma$
(not shown in the figure).
 Interestingly, as shown in Fig.\,\ref{Slw-Gg} (b), we find that
with increasing dot-resonator interaction,
the coherent signal for weak interdot Coulomb interaction is not only
shifted, but the step also turns into a dip.
This is consistent with the coherent signal of the Rabi frequency in the double dot
in the absence of the resonator showing
a dip and peak in the auto- and cross-correlation noise
spectra, respectively. \cite{Luo07085325,Jin1105}
It arises from the recovered symmetrical transition
tunneling channels [Fig.\,\ref{transition} (b) and (d)].
In the low-frequency regime, $\omega<|\omega_r|$,
the auto-correlation noise spectra of left and right lead
satisfy $S_L(\omega)=S_R(\omega)$. This is no longer true in the high-frequency regime  $\omega\gtrsim |\omega_r|$
for strong interdot Coulomb interaction, as will be shown in the following subsection.

\subsection{Regime close to the resonator frequency}\label{subsec_closetoresonatorfrequency} 

\begin{figure}
\centering
\centerline{\includegraphics*[width=1.03\columnwidth,angle=0]{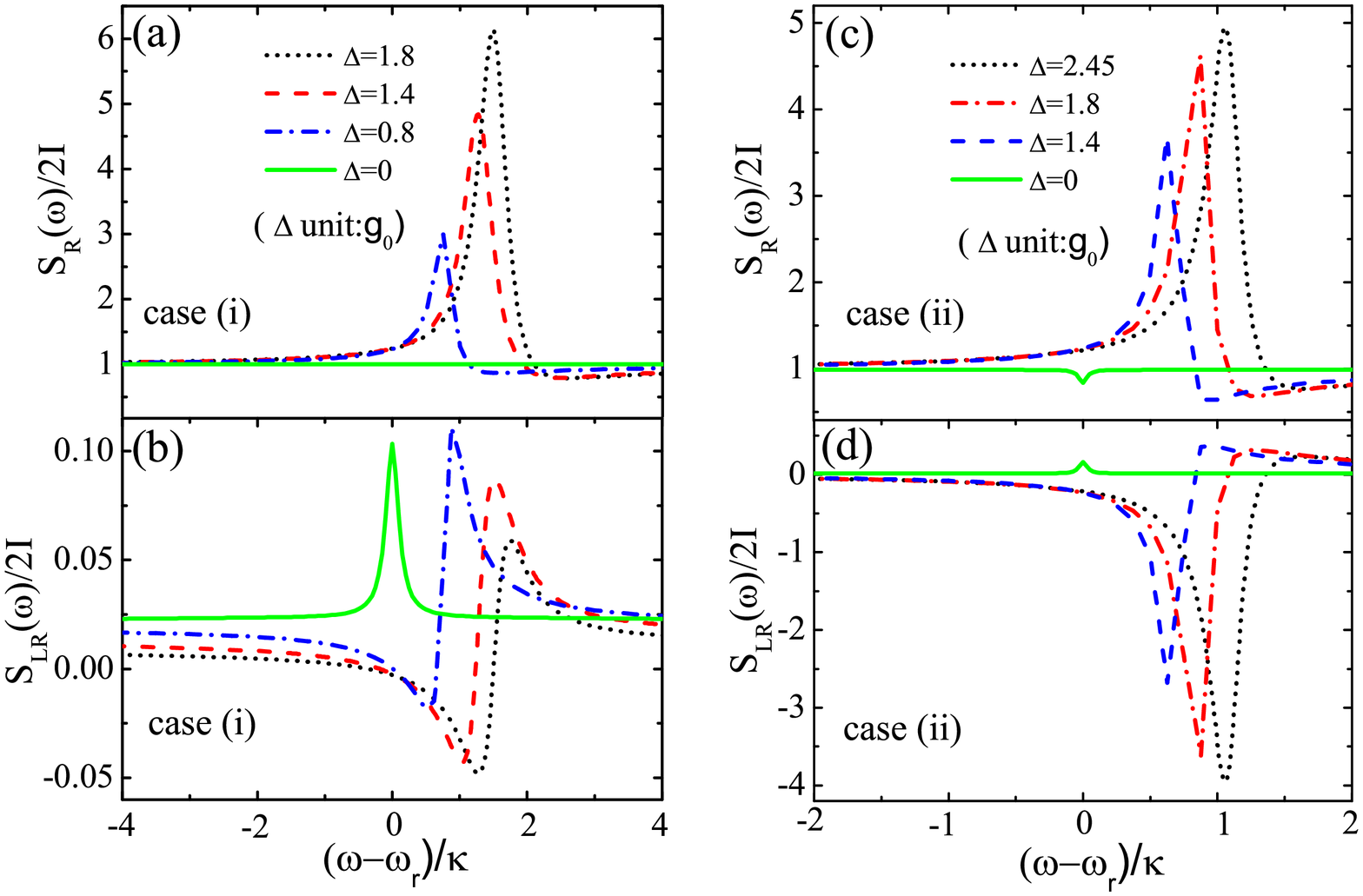}}
\caption{ (Color online)
The finite-frequency noise spectra for strong Coulomb 
interaction in the lasing regime. Panel (a) and (b) for 
strong interdot Coulomb interaction, and (c) and (b) 
for weak interdot Coulomb interaction.
The other parameters are the same as in Fig.\,\ref{stationary}.
} \label{Sw-Uinf}
\end{figure}

Before addressing the noise spectrum in the high-frequency regime,
let us briefly discuss its properties in the absence of the resonator.
It has been demonstrated \cite{Agu04206601,Luo07085325}
that the signal of the intrinsic Rabi frequency $\omega_0$ of the double dots
can be extracted
from the noise spectra. For instance, the
auto-correlation noise spectrum shows
a dip-peak structure and a dip at $\omega=\omega_0$ for strong
and weak interdot Coulomb interaction, respectively. \cite{Agu04206601,Luo07085325}
Considering the present parameter regime, where lasing is
induced for $\omega_0\approx\omega_r$
with very weak incoherent tunneling, $\Gamma=10^{-3}\omega_0$,
we find in the strong Coulomb interaction case
nearly Poissonian noise in the full-frequency regime,
with a small correction due to a weak coherent Rabi signal, i.e.,
$S_\alpha(\omega_0)/2I\sim1\pm 5\times 10^{-5}$.
The correction can be neglected compared to
the signal induced by the coupled resonator
as shown in Fig.\,\ref{Sw-Uinf}.


The signals in the high-frequency noise spectrum arise because of
transitions with the energy
$ E_{\pm,n}-E_{\pm,n-1}\approx \omega_r$.
They depend on the detuning in the same way
as the spectrum of the oscillator. \cite{And10053802}
Namely for positive detuning we find a signal at frequencies somewhat
higher than $\omega_r$ and for negative detuning at frequencies below $\omega_r$.

In contrast to the low-frequency case, for high frequencies
the spectra of the current in the left and
right junction, $S_L(\omega)$ and $S_R(\omega)$, do not have to be identical
due to the overall symmetry of the circuit broken by the resonator.
This feature has been demonstrated by the previous studies in
Refs.\,\onlinecite{Arm04165315,Har10104514}
for investigating the spectral properties of a resonator coupled to a
 SET and a SSET, respectively,
in non-lasing regime.
For the present studied setup in lasing regime, in this case we find significant differences
between the cases (i) and (ii) of strong and weak Coulomb interaction,
as illustrated in the left and right columns of Fig.\ref{Sw-Uinf}, respectively.
For strong Coulomb interaction the correlators are
\begin{eqnarray}
 \langle I_L(t) I_L(0)\rangle &=& \sum_n\langle n| \langle 0| \rho_{I_L}(t) |0\rangle |n\rangle\nonumber\\
 \langle I_R(t) I_R(0)\rangle &=& \sum_n\langle n| \langle R| \rho_{I_R}(t) |R\rangle |n\rangle \, ,
\end{eqnarray}
while for weak Coulomb interaction we have
\begin{eqnarray}
 \langle I_L(t) I_L(0)\rangle &=&\sum_n\langle n|\left[
                                 \langle 0| \rho_{I_L}(t) |0\rangle+\langle R| \rho_{I_L}(t) |R\rangle
                                 \right]  |n\rangle \nonumber\\
 \langle I_R(t) I_R(0)\rangle &=& \sum_n\langle n|\left[
                                  \langle R| \rho_{I_R}(t) |R\rangle+\langle 2| \rho_{I_R}(t) |2\rangle
                                   \right] |n\rangle \, .
                                   \nonumber\\
\end{eqnarray}
Here we introduced the density matrix $\rho_{I_i}(t)$ which
satisfies the master equation (\ref{UME})
with the initial condition $\rho_{I_i}(0)=\hat I_i \rho^{\rm st}$ ($i=L,R$).

For strong Coulomb interaction only $S_R(\omega)$ couples directly to the
state $|R\rangle$, which in turn couples resonantly to the oscillator. 
As a result we observe the signal at $\omega\approx\omega_r$ only in $S_R(\omega)$, while $S_L(\omega)\approx 1$ is unaffected by the oscillator.
In contrast, in case (ii), where we allow the state $|2\rangle$ to participate,
we again find a symmetry between the currents through the right and left junction
and $S_L(\omega)=S_R(\omega)$, as well as the anti-symmetry between
the auto- and cross-correlation noise spectrums,
i.e., roughly $S_{\alpha}(\omega)/2I\approx1+\Delta S(\omega)$ and
$S_{LR}(\omega)/2I\approx-\Delta S(\omega)$ with the signal $\Delta S(\omega)$
changing sign leading to a peak and dip as function of frequency.
Furthermore, in contrast to the low-frequency regime, the noise spectrum
at high frequency shows a Fano-resonance profile. It displays the same mechanism
as observed by Rodrigues \cite{Rod09067202} that the detector (here the double quantum
dot) feels the force in two ways, namely the original one from
the voltage-driven tunneling and the one from the resonator.
It arises from a destructive inference between
the two transition paths between $|g\ra$ and $|e\ra$, i.e,
a direct tunneling channel through the leads and
a transition assisted by the absorption and emission at the detection
 frequency.
Still, we like to mention that the present Fano-resonance profile occurs in the
current noise spectra differs from the result presented in Ref.\,\onlinecite{Rod09067202}
where the resonator coupled to SET showed the Fano-resonance in the SET charge noise spectra.


\section{summary and discussion}
\label{thsum}
We have evaluated the frequency-dependent noise spectrum
of the transport current through a coupled dot-resonator system in the
lasing regime, in a situation when incoherent tunneling induces a population inversion.
We considered both
strong and weak interdot Coulomb interactions, in the latter case
taking into account the doubly occupied state as well.
Both situations lead to a similar behavior of the zero-frequency shot noise but
to different features in the finite-frequency noise spectrum.

When the system approaches the lasing regime
the zero-frequency shot noise is enhanced strongly showing a remarkable
super-Poissonian distribution.
When the resonator is in the lasing state, the shot noise displays
sub-Poissonian characteristics. The current follows here the behaviour of the
photon distribution, which is also super-Poissonian as one approaches the lasing regime and becomes
sub-Poissonian near resonance.

We  found that the average photon number
and the corresponding Fano factor, as well as the average current and its Fano factor
 in the lasing regime is larger
for weak interdot Coulomb interaction than for
strong interaction.
The zero-frequency shot noise
could be  detected with current experimental technologies. E.g.,
a quantum point contact coupled to the dots has been demonstrated to detect in
 real-time  single electron tunneling
through the double dot. \cite{Gus06076605,Fuj061634}

Considering the finite-frequency noise spectra
we found pronounced characteristic structures
in the low- and high-frequency regimes reflecting
the  coherent dynamics of the coupled dot-resonator system.
At low but finite frequencies the coherent dynamics
of the oscillator leads to a peak at the eigen Rabi frequency of the coupled system.
At frequencies close to that of the resonator, due to the excitations of the photons in the
 resonator,
 we found for strong interdot Coulomb interaction a strongly asymmetric signal
in the auto-correlation noise spectra of the left and right junction. Symmetry is restored for
weak interdot Coulomb interaction. The difference arises from the asymmetrical and symmetrical
incoherent tunneling channels induced by strong and weak interdot Coulomb interactions, respectively.

\acknowledgments
J. S. Jin acknowledges support from
the National Natural Science Foundation of China NSFC (10904029 and 11274085)
as well as the support by the Ministry of Science, Research and the Arts of the State of Baden-W\"urttemberg.

\end{document}